# Particle growing mechanisms in Ag-ZrO$_2$ and Au-ZrO$_2$ granular films obtained by pulsed laser deposition

Zorica Konstantinović[1], Montserrat García del Muro[1], Manuel Varela[2], Xavier Batlle[1] and Amílcar Labarta[1]

[1]Departament de Física Fonamental and Institut de Nanociència i Nanotecnologia, Universitat de Barcelona, Martí i Franquès 1, 08028 Barcelona, Spain
[2]Departament de Física Aplicada i Òptica and Institut de Nanociència i Nanotecnologia, Universitat de Barcelona, Martí i Franquès 1, 08028 Barcelona, Spain

**Abstract**
Thin films consisting of Ag and Au nanoparticles embedded in amorphous ZrO$_2$ matrix were grown by pulsed laser deposition in a wide range of metal volume concentrations in the dielectric regime (0.08<$x_{Ag}$<0.28 and 0.08<$x_{Au}$<0.52). High resolution transmission electron microscopy (TEM) showed regular distribution of spherical Au and Ag nanoparticles having very sharp interfaces with the amorphous matrix. Mean particle size determined from X-ray diffraction agreed with direct TEM observation. The silver mean diameter increases more abruptly with metal volume content than that corresponding to gold particles prepared under the same conditions. Two mechanisms of particle growing are observed: nucleation and particle coalescence, their relative significance being different in both granular systems, which yields very different values of the percolation threshold ($x_c$(Ag)~0.28 and $x_c$(Au)~0.52).
Nanoparticles 61.46.Df,
Transmission electron microscopy 68.37.Lp,
Thin film structure and morphology 68.55.-a

## 1. Introduction

Granular insulating films, constituted of metallic particles embedded into a dielectric matrix, comprise a very active research field due to their relevant basic properties and potential applications [1]. When particles are reduced to nanometers size their properties are different from those of the bulk state, arising a wide variety of new phenomena, such as finite-size and surface effects, interparticle interactions, and enhanced properties [2, 3]. In particular, from a fundamental point of view, these composite systems show a variety of behaviors related to percolation processes that the standard percolation theories have not satisfactorily explained yet [4, 5, 6]. From technological aspect, spherical particles of noble metals homogeneously dispersed in dielectric matrix exhibit promising optical applications [7, 8, 9], associated with its large third order non-linear susceptibility [10, 11, 12] and ultrafast response [13].

In order to correlate properly the observed behavior with the corresponding microstructure and compare with theoretical predictions, the experimental model system should contain a narrow size distribution of immiscible nanoparticles very well defined with respect to the matrix. Recent works [14, 15, 16, 17] demonstrate that pulsed laser deposition (PLD) produces granular thin films which are closer to this nanostructure model than those films obtained by other techniques, such as sputtering [18, 19, 20, 21], sol-gel methods [22, 23] or ion implantation [24, 25]. In particular, granular thin films of composition Co-ZrO$_2$ [14, 17], Ag-Al$_2$O$_3$ and Au-Al$_2$O$_3$ [15, 16] have been previously grown by PLD and studied.

In this paper, we present a complete study of the preparation and structural characterization of granular Ag-ZrO$_2$ and Au-ZrO$_2$ thin films grown by PLD from a single composite target within a wide range of volume fraction x of the noble metal in the dielectric regime (0.08<$x_{Ag}$<0.28 and 0.08<$x_{Au}$<0.52). The structural results are compared aiming to stress the effect of the actual microstructure on the percolation threshold. In particular, we observe that for both noble metals the mean size of the particles increases with the metal content in a similar manner to that reported for samples with Al$_2$O$_3$ matrix using a two target technique [15,16]. However, Ag-ZrO$_2$ and Au-ZrO$_2$ granular films show two different mechanisms of particle growing as a function of the metal content, which yields very different values of the physical percolation threshold.

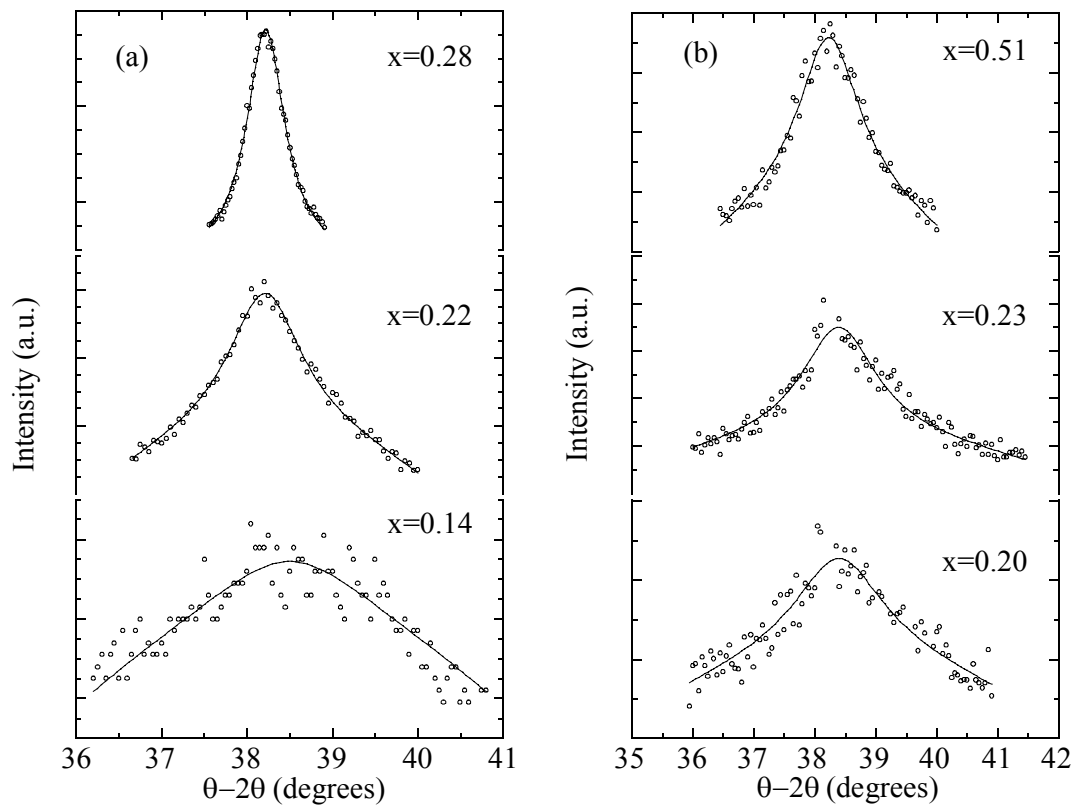

**Figure 1.** (111) X-ray diffraction for (a) Ag-ZrO$_2$ and (b) Au-ZrO$_2$ for different metal content. Solid lines are fits to a pseudo-Voigt function.

## 2. Experimental details

Ag-ZrO$_2$ and Au-ZrO$_2$ granular films were grown by KrF laser ablation (wavelength of 248 nm, pulse duration of $\tau$=34 ns). The samples were deposited at room temperature in a vacuum chamber with rotating composite targets made of sectors of ZrO$_2$ and noble metal (silver or gold). Several surface ratios of target components led to obtain samples with different volume fractions x of Ag/Au, ranging from metallic to dielectric regimes. The distance between target and substrate was fixed to 35 mm. The laser fluency typically used was about 2 J/cm$^2$. Zirconia was stabilized with 7 mol.% Y$_2$O$_3$, which provides the matrix with very good properties, such as good oxidation resistance, thermal expansion coefficient matching that of metal alloys and very high fracture toughness values. It has been observed that ZrO$_2$ matrix seems to better coat metallic nanoparticles [14], enabling the occurrence of sharper interfaces between the amorphous matrix and nanoparticles. Besides, the high oxygen affinity of ZrO$_2$ prevents oxidation of the metallic nanoparticles.

Average sample composition was determined by microprobe analyses. The size distribution of noble metal nanoparticles were determined from transmission electron microscopy and compared with the average particle size estimated from X-ray diffraction (XRD). The thickness of the films prepared for XRD were within 200 and 300 nm. The substrates for TEM experiments were membrane windows of silicon nitride, which enabled direct observation of as deposited samples. Samples prepared for TEM observations were quite thin (in the order of a few nm) and could be considered as a result of almost 2D growing processes. However, this is a necessary condition to achieve visualization of the nanoparticles by this technique. Besides, thin films of a few nanometers in thickness prepared by laser ablation do not grow layer by layer in a continuous manner, but they show inhomogeneous nucleation processes around islands of a hundred nanometers, which yields a quasi-three dimensional effective growing of the embedded granular structures. Consequently, TEM observations are always two-dimensional images which in deed can be used to detect the existence of three-dimensional percolative structures in the sample.

## 3. Results and discussion

Figure 1 shows the X-ray $\theta$-2$\theta$ diffraction patterns around the (111) reflection of Ag (a) and Au (b) measured for Ag-ZrO$_2$ and Au-ZrO$_2$ films, respectively, for various volume concentrations. The peaks were well defined for high volume content of metal, indicating good crystallinity of the noble nanoparticles.

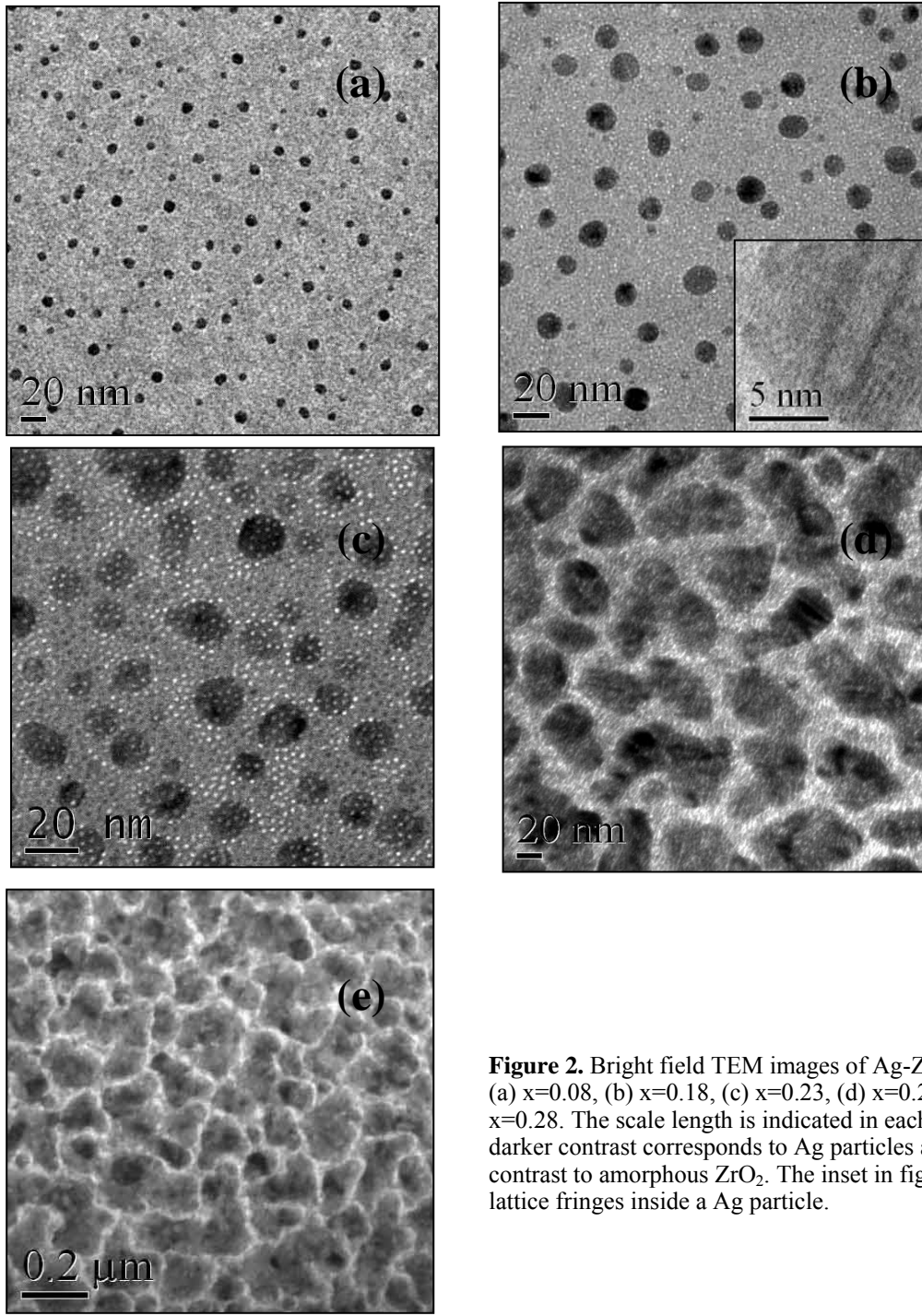

**Figure 2.** Bright field TEM images of Ag-ZrO$_2$ films with (a) x=0.08, (b) x=0.18, (c) x=0.23, (d) x=0.26 and (e) x=0.28. The scale length is indicated in each image. The darker contrast corresponds to Ag particles and the lighter contrast to amorphous ZrO$_2$. The inset in figure 2(b) shows lattice fringes inside a Ag particle.

We should note that, at low volume concentration (x=0.14 for Ag and x=0.20 for Au), peaks could no longer be distinguished in conventional θ-2θ scan, but they were obtained by XRD with grazing beam incidence (ω=0.30°) in order to maximize the diffraction intensity. For samples with even smaller volume concentrations (x<0.14) diffraction peaks could not be distinguished even with grazing beam incidence because of too low metal concentration.

From the line broadening it was possible to determine the average particle size by fitting the diffraction peaks to a pseudo-Voigt function (solid lines in figure 1). In table 1, we show the obtained peak positions 2θ$_B$ and the full width at half maximum (FWHM). These two parameters are used to determine average values of the particle diameter D$_M$ via Scherrer's formula ($D_M = 0.94 \frac{\lambda}{FWHM} \frac{1}{\cos(\theta_B)}$ with λ=1.54 Å) as a function of metal volume concentration. The peak position (2θ$_B$) remains almost constant for all x, this value being similar in both Ag and Au diffraction peaks and corresponding to (111) reflection of fcc crystalline structure.

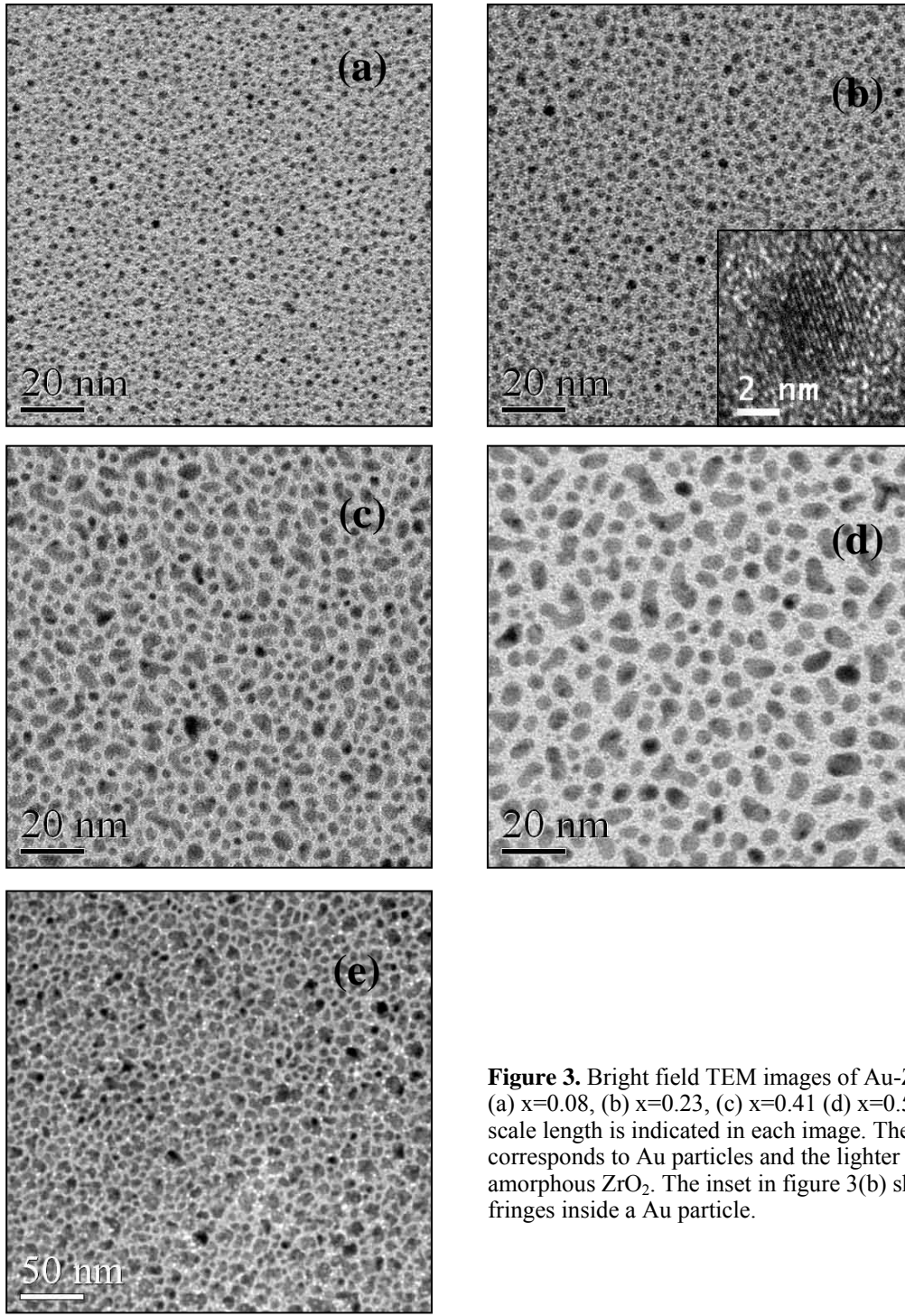

**Figure 3.** Bright field TEM images of Au-ZrO$_2$ films with (a) x=0.08, (b) x=0.23, (c) x=0.41 (d) x=0.50 (e) x=0.52. The scale length is indicated in each image. The darker contrast corresponds to Au particles and the lighter contrast to amorphous ZrO$_2$. The inset in figure 3(b) shows lattice fringes inside a Au particle.

The average nanoparticle diameter increased with x in both granular films (table 1). Varying the metal content, we obtained average diameters of silver nanoparticles within 3 nm and 16 nm (0.14<$x_{Ag}$<0.28) and for gold within 5 nm and 7 nm (0.20<$x_{Ag}$<0.51). It is worth noting that under the same conditions of preparation Ag nanoparticles were obtained in a wider range of sizes than Au ones.

Although a global estimation of average nanoparticle diameters can be gained from XRD, TEM images enable direct observation of the nanoparticles even for very low metal concentrations. Typical TEM images are shown in figure 2 for Ag-ZrO$_2$ and in figure 3 for Au-ZrO$_2$. The dark regions correspond to the Ag and Au particles and the light regions to the amorphous ZrO$_2$ matrix. The particles are seen to have clearly defined interfaces with the matrix. Moreover, the silver nanoparticles shown by TEM images appear to be significantly bigger and consequently, farther apart one to each other than gold particles for similar volume concentrations.

**Table 1.** Peak position $2\theta_B$, FWHM and average particle diameter $D_M$ for several values of x.

| $x_{Ag}$ | $2\theta_B$ (°) | FWHM (°) | $D_M$ (nm) |
|---|---|---|---|
| 0.28 | 38.22 | 0.54 | 16.2 |
| 0.22 | 38.22 | 1.04 | 8.4 |
| 0.14 | 38.50 | 2.41 | 3.6 |
| $x_{Au}$ | $2\theta_B$ (°) | FWHM (°) | $D_M$ (nm) |
| 0.51 | 38.23 | 1.20 | 7.3 |
| 0.23 | 38.40 | 1.51 | 5.8 |
| 0.20 | 38.40 | 1.81 | 4.9 |

**Table 2.** Most probable particle diameter $D_0$, average particle diameter $D_M$ ($=D_0\exp(\sigma^2/2)$) and distribution width $\sigma$ obtained from TEM data as a function of the volume concentration of noble metal.

| $x_{Ag}$ | $D_0$ (nm) | $D_M$ | $\sigma$ |
|---|---|---|---|
| 0.28 | 220 | 249.3 | 0.50 |
| 0.26 | 39 | 39.9 | 0.21 |
| 0.23 | 17 | 17.4 | 0.22 |
| 0.18 | 11 | 11.2 | 0.20 |
| 0.08 | 1.7 | 1.8 | 0.40 |
| $x_{Au}$ | $D_0$ (nm) | $D_M$ | $\sigma$ |
| 0.52 | 9.5 | 10 | 0.32 |
| 0.50 | 5.9 | 6.1 | 0.25 |
| 0.41 | 3.0 | 3.3 | 0.45 |
| 0.23 | 2.0 | 2.2 | 0.40 |
| 0.08 | 1.2 | 1.3 | 0.45 |

The insets in figures 2(b) and 3(b) show a magnification of selected silver and gold particles, respectively. The lattice fringes observed in the metal grains correspond to Ag/Au atomic planes indicating good crystallinity even for very low metal content. Lattice fringes are not present in the $ZrO_2$ matrix, in agreement with its amorphous nature.

At metal concentration, $x_{Ag}$<0.18 and $x_{Au}$<0.41, the particles have spherical shape (figure 2(a)-(b), figure 3(a)-(b)). For $x_{Ag}$>0.18 and $x_{Au}$>0.41, the neighboring particles start to coalesce, giving rise to larger particles not always with spherical shape (see elongated particles in figures 2(b)-(d) and 3(c)-(d)). With further increasing of the metal content, $x_{Ag}\geq0.28$ and $x_{Au}\geq0.52$, the particles form big aggregates (figure 2(e) and 3(e), note the different scale length in comparison with the rest of TEM images), indicating rapid approaching to the percolation threshold, above which metal forms a continuum.

TEM micrographs were analyzed in order to quantify the particle size distributions shown in figure 4 and 5 for Ag-$ZrO_2$ and Au-$ZrO_2$, respectively. The distributions of particle size are well described by a log-normal function:

$$f(D) = \frac{1}{\sqrt{2\pi}\sigma D}\exp\left[-\frac{\ln^2(D/D_0)}{2\sigma^2}\right]$$

where the fitting parameters $D_0$ and $\sigma$ are the most probable particle size and the width of the distribution, respectively (see table 2). At low Ag content, the particle size distribution is centered between 1 and 2 nm (figure 2(a) and 4(a)). Increasing the Ag content, the size distribution shifts to larger sizes, due to coalescence of smaller particles into the big ones (figure 4(b)-(d)), which produces a net narrowing effect on the particle size distribution. About $x_{Ag}$=0.28 (figure 4(e)) the size distribution broadens abruptly because of massive coalescence of the nanoparticles taking place at percolation. However, a quite different evolution of the microstructure is observed for Au-$ZrO_2$ as the Au content is increased. At low Au content, the width of the particle size distribution is similar to that observed for silver with $x_{Ag}$=0.08. Nevertheless, in this case, a very smooth shift of the size distribution towards larger sizes is observed even for Au contents as high as $x_{Au}$=0.41 (figures 5(a)-(c)), suggesting that in a wide range of concentrations Au particles tend to be coated by the matrix, which minimizes particle coalescence and maintains the width of the size distribution almost constant. The onset of coalescence processes takes place about $x_{Au}$>0.41 (figure 5(d)) giving rise to a similar narrowing of the size distribution as it is also observed in Ag-$ZrO_2$, but in this case occurring at metal contents very close to percolation. Finally, at $x_{Au}$~0.52 massive coalescence of the nanoparticles arising from percolative processes takes place, which produces a broadening of the size distribution (figures 3(e) and 5(e)), as it is also observed for Ag-$ZrO_2$.

The dependence on the metal volume concentration of the average particle sizes $D_M$ for silver and gold, obtained from log-normal distributions (open symbols) and the broadening of (111) metal reflection (solid symbols), are in good agreement, as it is shown in figure 6. The values of the average particle size for both silver (triangles) and gold (circles) increase with noble metal concentration, but following very different behaviors. With increasing Au content, mean particle size slightly increases (figure 5(b)), since in this case and below about $x_{Au}$=0.4 particles grow essentially by condensation of the gold atoms available in the neighborhood of each nucleating seed (figures 5(a)-5(b)), according to TEM images (figures 3(a)-(b)). However, for Ag-$ZrO_2$, the mean particle size increases abruptly with $x_{Ag}$ because particle growth is arising from nucleation and further coalescence of neighboring particles even at low metal contents.

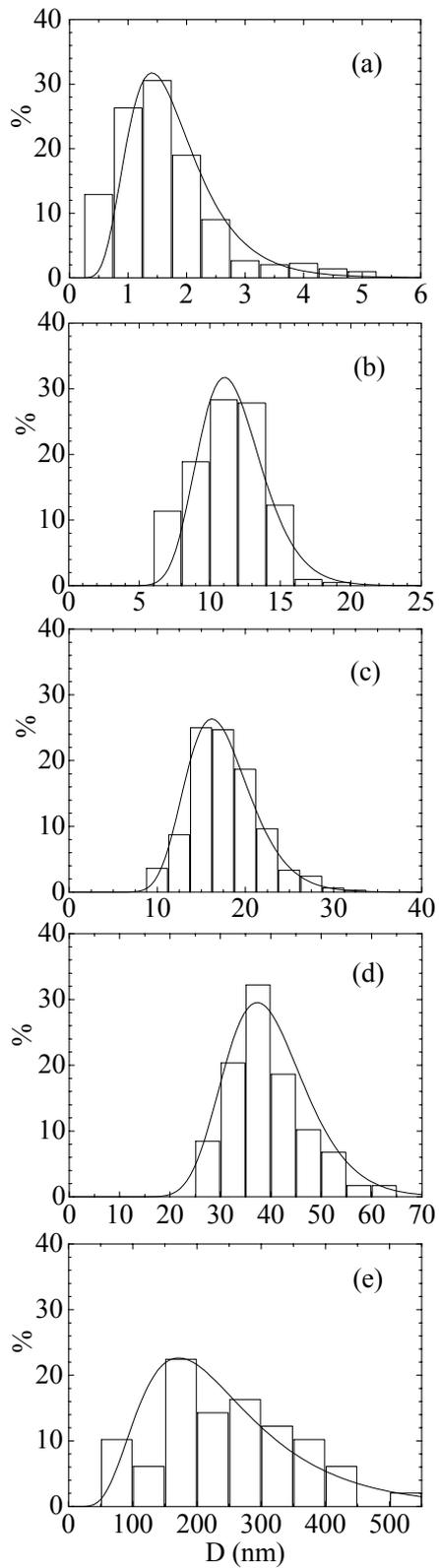

**Figure 4.** Distribution of particle sizes in percentage, obtained after analysis of the TEM micrographs for Ag-ZrO$_2$ with (a) x=0.08 (b) x=0.18 (c) x=0.23 (d) x=0.26 and (e) x=0.28. The solid lines are fits to a log-normal distribution

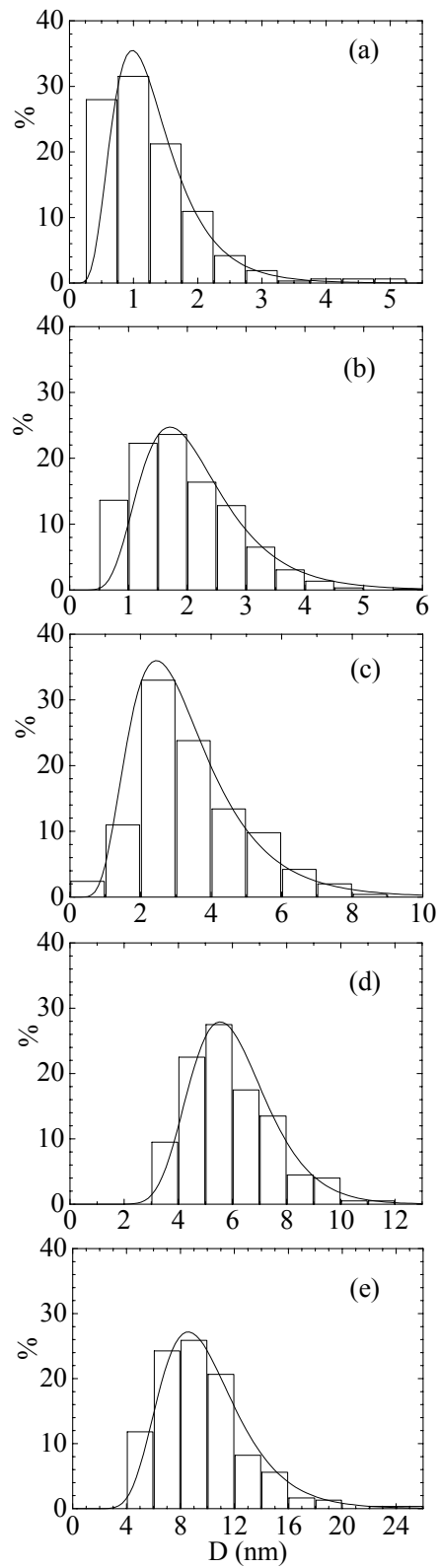

**Figure 5.** Distribution of particle sizes in percentage, obtained after analysis of the TEM micrographs for Au-ZrO$_2$ with (a) x=0.08 (b) x=0.22 (c) x=0.41 (d) x=0.50 and (e) x=0.52. The solid lines are fits to a log-normal distribution.

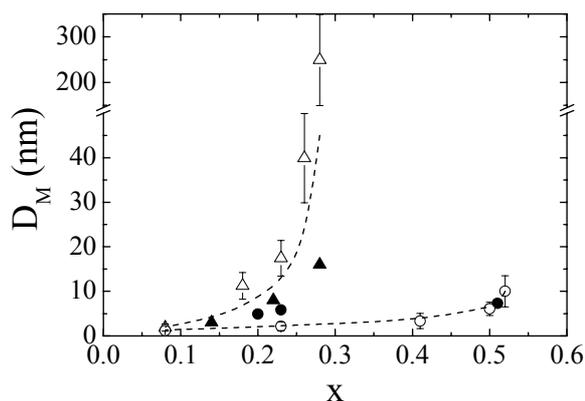

**Figure 6.** Mean size diameters of noble nanoparticles determined respectively from X-ray data (solid symbols) and log-normal distributions (open symbols) for Ag-ZrO$_2$ (triangles) and Au-ZrO$_2$ (circles). Error bars denote ± one standard deviation. Dashed lines are a guide for the eye.

The two mechanisms of particle growing observed in Ag-ZrO$_2$ and Au-ZrO$_2$ granular films give rise to very different values of the percolation threshold [26]. For Ag-ZrO$_2$, percolation threshold deduced from TEM images ($x_c$(Ag)~0.28) is very close to the theoretical prediction for the model of random percolation of hard spheres [27]. In contrast, particle coalescence in Au-ZrO$_2$ is inhibited by the better efficiency of the matrix to coat gold particles with respect to silver ones, which retards the occurrence of percolating processes, shifting the critical value of the metal content to $x_c$(Au)~0.52 (see figure 3(e)).

The diameters of gold nanoparticles for Au-ZrO$_2$ are similar to those reported for Au-Al$_2$O$_3$ granular films for similar metal contents [2,16], suggesting that the nature of the matrix does not play a crucial role in this case, provided it is amorphous. In contrast, spherical nanoparticles in Ag-ZrO$_2$ are observed in a wider range of diameters (within 2 and 17 nm) than that corresponding to Ag-Al$_2$O$_3$ (within 1 and 3 nm) [15].

## 5. Conclusions

We have demonstrated that pulsed laser deposition is an appropriate technique to prepare silver and gold nanoparticles embedded in ZrO$_2$ matrix, in a wide range of volume concentration (0.08<$x_{Ag}$<0.28 and 0.08<$x_{Au}$<0.52). As prepared films without ulterior thermal treatment show rounded particles of noble metal, which are crystalline and have sharp interfaces with the matrix. The mean nanoparticle diameter increases with metal volume concentration. Silver nanoparticles are obtained in a wider range of diameters (1-200 nm) than that corresponding to gold ones (1-10 nm) obtained under the same preparation conditions. Besides, Ag-ZrO$_2$ and Au-ZrO$_2$ granular films show very different microstructures, as a consequence of the relative contribution of the two particle-growing mechanisms: nucleation and particle coalescence. Consequently, the percolation thresholds are very different in these two systems, ($x_c$(Ag)~0.28 and $x_c$(Au)~0.52).

## Acknowledge

We would like to thank the staff of the scientific and technical facilities of UB. Financial support of the Spanish CICYT (MAT2003-01124) and the Catalan DURSI (2005SGR00969) are gratefully recognized. ZK thanks Spanish MEC for the financial support through Juan de la Cierva program.